\documentclass{elsart}
\usepackage{textcomp,latexsym,amssymb}
\usepackage[dvips]{graphicx,epsfig}

\newcommand{\reaction}[4]{\textsuperscript{#1}#2 + \textsuperscript{#3}#4}
\newcommand{\iso}[2]{\textsuperscript{#1}#2}

\begin{document}
\begin{frontmatter}

\title{Investigation of the role of neutron transfer in the fusion of $^{32,34}$S 
with $^{197}$Au,$^{208}$Pb using quasi-elastic scattering}
\author[ANU,ikf]{T.J.~Schuck},
\author[adfa]{H.~Timmers},
\author[ANU]{M.~Dasgupta}
\address[ANU]{Department of Nuclear Physics,
Australian National University, Canberra, ACT 0200, Australia}
\address[ikf]{Institut f\"ur Kernphysik, Universit\"at Frankfurt, D-60486 Frankfurt, Germany}
\address[adfa]{School of Physics, University of New South Wales at the 
Australian Defence Force Academy, Canberra, ACT 2600, Australia}

\begin{abstract}

Excitation functions for quasi-elastic scattering have been 
measured at backward angles 
for the systems \reaction{32,34}{S}{197}{Au} and \reaction{32,34}{S}{208}{Pb} for energies
spanning the Coulomb barrier. 
Representative distributions, sensitive to the low energy part of the fusion barrier distribution, 
have been extracted from the data. 
For the fusion reactions of $^{32,34}$S with $^{197}$Au  
couplings related to the nuclear structure of $^{197}$Au appear to be dominant in shaping the low energy part
of the barrier distibution. For the system \reaction{32}{S}{208}{Pb} the barrier distribution is broader and 
extends further to lower energies, 
than in the case of \reaction{34}{S}{208}{Pb}. This is consistent with the interpretation that 
the neutron pick-up channels are energetically more favoured in the $^{32}$S induced reaction 
and therefore couple more strongly to the relative motion. 
It may also be due to the increased collectivity of $^{32}$S, 
when compared with $^{34}$S. 

\end{abstract}
\end{frontmatter}

\noindent
{\bf PACS codes}
heavy-ion nuclear reaction 25.70, heavy ion induced fusion 25.70J, nuclear 
scattering 25.30

\noindent
{\bf KEYWORDS}
NUCLEAR REACTIONS $^{197}$Au,$^{208}$Pb($^{32,34}$S,X), 
$E=115-175$\,MeV;
measured quasi-elastic scattering excitation functions;
deduced representations of fusion barrier distributions;
subbarrier fusion, channel-coupling, neutron transfer.

\section{Introduction}

Measured cross sections for heavy ion fusion at energies below the Coulomb barrier 
show strong isotopic dependences and exceed theoretical predictions 
based on a single barrier penetration model by several orders
of magnitude~\cite{reisdorf,nandarev}. This has been observed 
for a wide range of systems and is understood to arise from the coupling 
of the relative motion of the interacting nuclei 
to their rotational and vibrational
states or to particle transfer channels.
The coupling gives rise to a distribution of fusion barriers $D(E)$~\cite{dasso}.
Experimentally, a representation $D^{fus}(E)$ of this distribution can be extracted from 
precision measurements of fusion
excitation functions $\sigma^{fus}(E)$ using~\cite{rowley}:

\begin{equation}
D^{fus}(E) = \frac{d^2(E\sigma^{fus})}{dE^2}
\end{equation}

At energies below the Coulomb barrier quasi-elastic scattering 
excitation functions $d\sigma^{qel}$/${d\sigma^R}(E)$, measured at backward angles, 
have been found~\cite{qeldist} to be 
another suitable means to extract representations $D^{qel}(E)$  
of the distribution $D(E)$ with:

\begin{equation}
D^{qel}(E) = -\frac{d}{dE} \left(\frac{d\sigma^{qel}}{d\sigma^R}(E)\right)
\label{dqel}
\end{equation}

In this technique, which is generally less complex than detailed fusion measurements,
 quasi-elastic scattering is understood to 
comprise elastic and inelastic scattering and also particle transfer channels. 
Experiments which have employed this approach have recently been 
carried out by several other groups ~\cite{capurro,santra,sinha}. 

In previous work~\cite{qeldist} it has been clearly
demonstrated  that the quasi-elastic scattering representation $D^{qel}(E)$ is not identical to
the representation $D^{fus}(E)$ extracted from fusion data, 
although it appears that this has not 
been appreciated in all studies. 
In particular, $D^{qel}(E)$ has been shown  to be
insensitive to the high energy part of the barrier distribution $D(E)$.
The  quasi-elastic scattering representations are therefore most useful for investigating 
couplings, which produce signatures in the low energy part of $D(E)$. 
This is the case when the relative motion of the two nuclei couples to positive Q-value 
channels~\cite{nandarev}. 

Indeed, the comparison of the representations $D^{fus}(E)$ 
and $D^{qel}(E)$ for the systems
 \reaction{40}{Ca}{90,96}{Zr} has shown that the effect of positive 
Q-value neutron 
transfer channels on the fusion dynamics are clearly seen in the representation $D^{qel}(E)$~\cite{cazr}.
The low-lying collective states in the two Zr isotopes have very similar excitation energies 
and deformation parameters $\beta_2$, 
and the main differences between these two systems are in their 
Q-values for neutron transfer. In the heavier system the calcium nucleus can pick-up
as many as eight neutrons in transfer reactions with positive Q-value, 
whereas the equivalent channels in the lighter system
all have negative Q-values. This pronounced difference has been found to be 
reflected in both types of representations, $D^{fus}(E)$ and $D^{qel}(E)$, 
measured for these systems~\cite{cazr}. The straight-forward measurement of quasi-elastic scattering excitation functions 
thus appears to be a promising tool to investigate the role of positive Q-value transfer channels in fusion.

The fusion reactions of the sulphur projectiles $^{32,34}$S with $^{197}$Au and $^{208}$Pb
are a suitable test case for
this new experimental approach to the dynamics of fusion.
This is apparent from Table \ref{qvalues}, which 
shows the Q-values for the pick-up of one and two neutrons for these systems. Also shown are the equivalent 
Q-values for the 
reactions $^{36}$S + $^{197}$Au, $^{208}$Pb. It is apparent that with decreasing projectile mass 
the Q-values progessively favour the neutron pick-up channels. The $^{32}$S and $^{34}$S projectile nuclei 
have similar structure, with the lowest $2^+$ states not being 
very different in terms of excitation energy and deformation parameter (see Table \ref{deform}).

\begin{table}
\begin{center}
\begin{tabular}{cccc}
\hline
& \reaction{32}{S}{197}{Au} & \reaction{34}{S}{197}{Au} & 
\reaction{36}{S}{197}{Au} \\ \hline 
 1n & +0.569 &$-$1.086 &$-$3.768 \\ 
 2n & +5.342 & +2.158&$-$2.377 \\ \hline \\ \hline
 & \reaction{32}{S}{208}{Pb} & \reaction{34}{S}{208}{Pb} & 
\reaction{36}{S}{208}{Pb} \\ \hline  
 1n &+1.274  &$-$0.382 &$-$3.064 \\ 
 2n & +5.953 &+2.769 &$-$1.766 \\ \hline 
\end{tabular}
\caption{The Q-values (in MeV) for the pick-up of one (1n) and two (2n) neutrons from the target nucleus for
 the systems studied in this work and the reactions \reaction{36}{S}{197}{Au},\iso{208}{Pb}.}
\label{qvalues}
\end{center}
\end{table}

The systems $^{32,34}$S + $^{197}$Au, $^{208}$Pb
are intermediate in mass between a large number of lighter fusion reactions, which have been
well-studied using the coupled-channels framework~\cite{nandarev}, and the more massive 
systems employed for the
synthesis of super-heavy elements~\cite{hofmann}. 
Results may thus 
indicate, if the representation $D^{qel}(E)$ is 
also applicable to this
important latter group of fusion reactions, for which multiple neutron transfer may lead to 
macroscopic effects such as neutron-flow or neck-formation.

\begin{table}
\begin{center}
\begin{tabular}{ccc}
\hline
    & $E_2$ [MeV] & $\beta_2$ \\ \hline
\iso{32}{S} & 2.230 & 0.31 \\
\iso{34}{S} & 2.127 & 0.25 \\ 
\hline 
\end{tabular}
\caption{Excitation energies $E_2$ and deformation parameters $\beta_2$ for the first 2$^+$ states 
of \iso{32}{S} and \iso{34}{S}.}
\label{deform}
\end{center}
\end{table}

This paper presents detailed 
measurements at backward angles of quasi-elastic scattering excitation functions for 
the two pairs of reactions \reaction{32,34}{S}{197}{Au} and 
\reaction{32,34}{S}{208}{Pb}, from which representations $D^{qel}(E)$ of the fusion barrier distribution have
been extracted. 

\section{Experimental Method}

The experiments were performed with \iso{32,34}{S}-beams from the 14UD Pelletron accelerator at the 
Australian National University in the energy range $E_{lab}= $ 90.0--180.0\,MeV\@.
Three different self-supporting Au targets were used, with
thicknesses in the range 140--170\,$\mu$g/cm$^2$. The \iso{208}{PbS} target was 
140\,$\mu$g/cm$^2$ thick, evaporated onto a 
$\sim$ 20\,$\mu$g/cm$^2$ carbon backing.
 The target thickness was determined by measuring the energy loss of elastically scattered projectiles 
in the target at a backward angle.

\begin{figure}
\begin{center}
\includegraphics{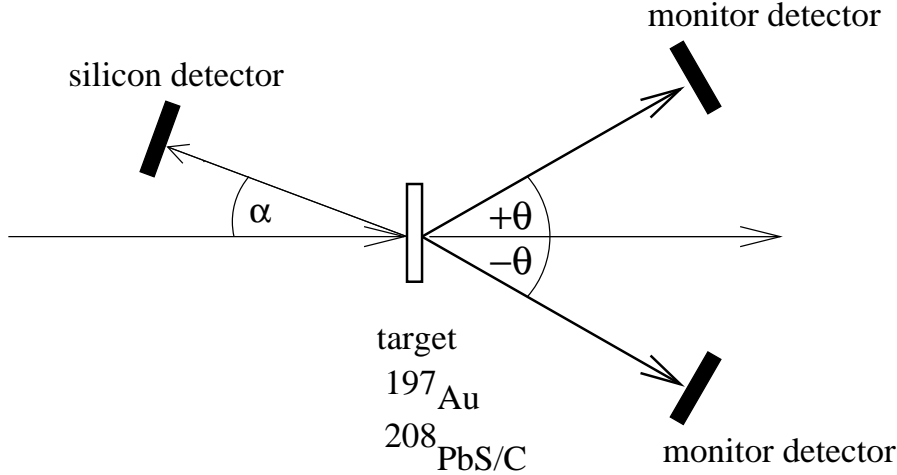}
\caption{A schematic diagram of the experimental setup used. The backward silicon surface barrier detector was at an angle 
$\alpha=21^\circ$. In the study of \reaction{32}{S}{208}{Pb} a $\Delta E-E_{res}$ telescope detector at $\alpha=10^\circ$
was used instead. The beam was monitored using two silicon detectors at $\theta = \pm 30^\circ$.}
\label{setup}
\end{center}
\end{figure}

A schematic diagram of the experimental setup is shown in Figure~\ref{setup}. 
In the initial study of the \reaction{32}{S}{208}{Pb} reaction~\cite{heikophd} 
quasi-elastic scattering was detected 
at a scattering angle of $\theta_{lab}= 170^\circ$. 
An energy loss signal $\Delta E$ 
was measured with a gas ionisation detector.
The gas detector was backed by a silicon surface barrier detector 
which detected the residual energy 
$E_{res}$ of the scattered nuclei. The $\Delta E$ signal allowed the separation of 
the charged particle transfer contributions to the quasi-elastic scattering yield.
Since this separation was not required for the interpretation of the data, 
in the other experiments the $(\Delta E-E_{res})$ detector telescope was replaced with a single
silicon surface barrier detector at $\theta_{lab}= 159^\circ$. 

For normalisation purposes, two silicon surface barrier detectors
were placed at scattering angles $\theta_{lab}= \pm 30^\circ$
to measure Rutherford scattering of projectiles. 
For some preliminary measurements, instead of these two monitor detectors, a readily available gas-ionisation
detector~\cite{erdadet} at $\theta_{lab}=30.35^\circ$ was employed. Both setups gave consistent results, so that the
data have been combined.

\begin{figure}
\begin{center}
\includegraphics[width=\linewidth]{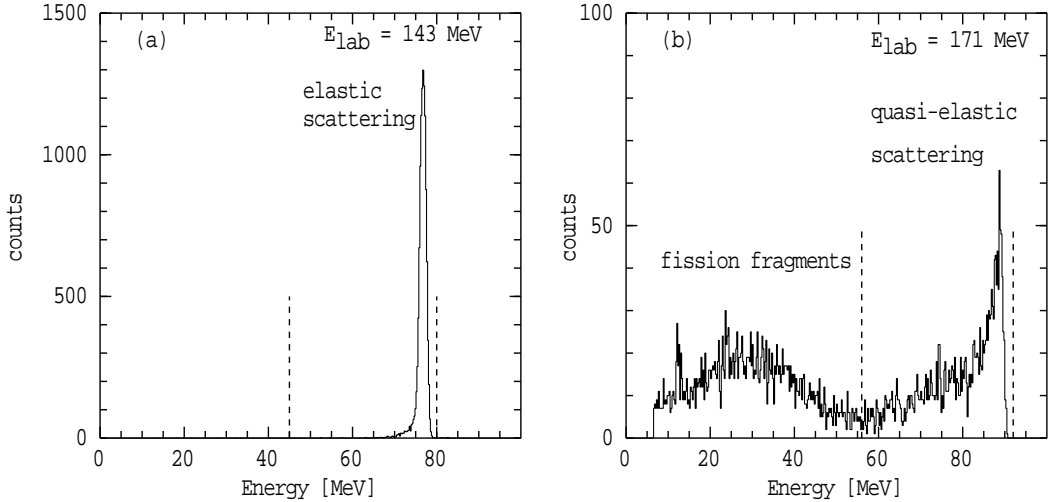}
\caption{Typical energy spectra from the backward silicon detector at $\theta_{lab} = 159^\circ$ for the system
\reaction{32}{S}{197}{Au} at
 (a) $E_{lab}=143$\,MeV and (b) $E_{lab}=171$\,MeV (b). 
At the low energy all scattering is elastic, in the high energy spectrum
fission fragments and quasi-elastic scattering can be identified.
The vertical, dashed lines indicate the gate which was used to integrate the quasi-elastic scattering yield.}
\label{sispec}
\end{center}
\end{figure}

Figure~\ref{sispec} shows typical 
energy spectra from the backward silicon detector for the system \reaction{32}{S}{197}{Au}. 
The spectra for the other three systems 
are similar. At low beam energies the spectrum only shows 
a well-defined peak of elastically scattered sulphur nuclei.
With increasing energy the peak gradually develops a low energy tail as the yield of non-elastic scattering events rises and 
elastic scattering is diminished.
At the higher energies fission fragments from quasi-fission and fission following compound nucleus formation
are also detected. The energy window chosen for the integration 
of the quasi-elastic scattering yield is indicated in the figure.

For all experiments the quasi-elastic scattering yield, measured at the backward angle, 
comprising the sum of elastic, inelastic and transfer events, 
was divided by the Rutherford scattering
yield detected at forward angles. These ratios have been normalised to unity at energies
well below the Coulomb barrier, where only
Rutherford scattering is observed. The normalised ratios thus represent
the quasi-elastic scattering excitation function 
$\frac{d\sigma^{qel}}{d\sigma^{R}}(E)$, where
the indices $qel$ and $R$ indicate quasi-elastic scattering and Rutherford scattering,
respectively. 

\begin{figure}
\begin{center}
\includegraphics{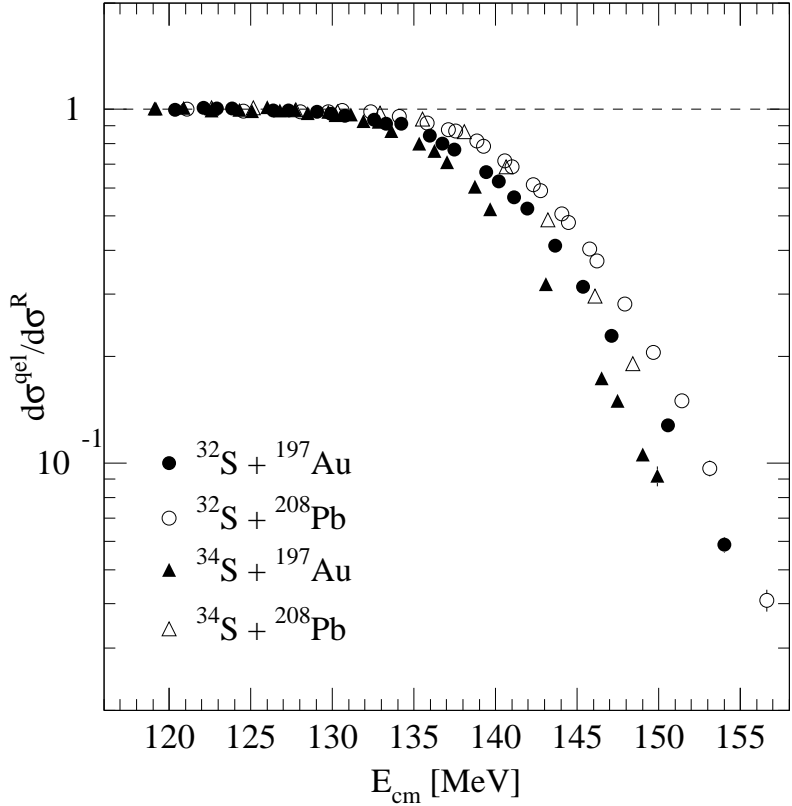}
\caption{The measured quasi-elastic scattering excitation functions.
Centrifugal energies and energy loss in the target have been subtracted to allow a direct comparison of the 
four measurements. Statistical uncertainties are smaller than the symbol size, unless indicated otherwise.}
\label{allcs}
\end{center}
\end{figure}

The measured quasi-elastic scattering excitation functions 
are shown in Figure \ref{allcs}. With the exception of the highest energies, 
where the quasi-elastic scattering yield is low, the statistical uncertainty is better than $1\%$. 
The energy scale has been adjusted for energy loss in
the target, and the centrifugal energy corresponding to the respective 
detection angle has been subtracted, as described in~\cite{qeldist}. 
The excitation functions decrease smoothly with energy. The apparent energy shifts between
the four sets of data reflect the expected differences in Coulomb barrier height. 

\section{Discussion of the Experimental Data}

\begin{figure}
\begin{center}
\includegraphics[width=\linewidth]{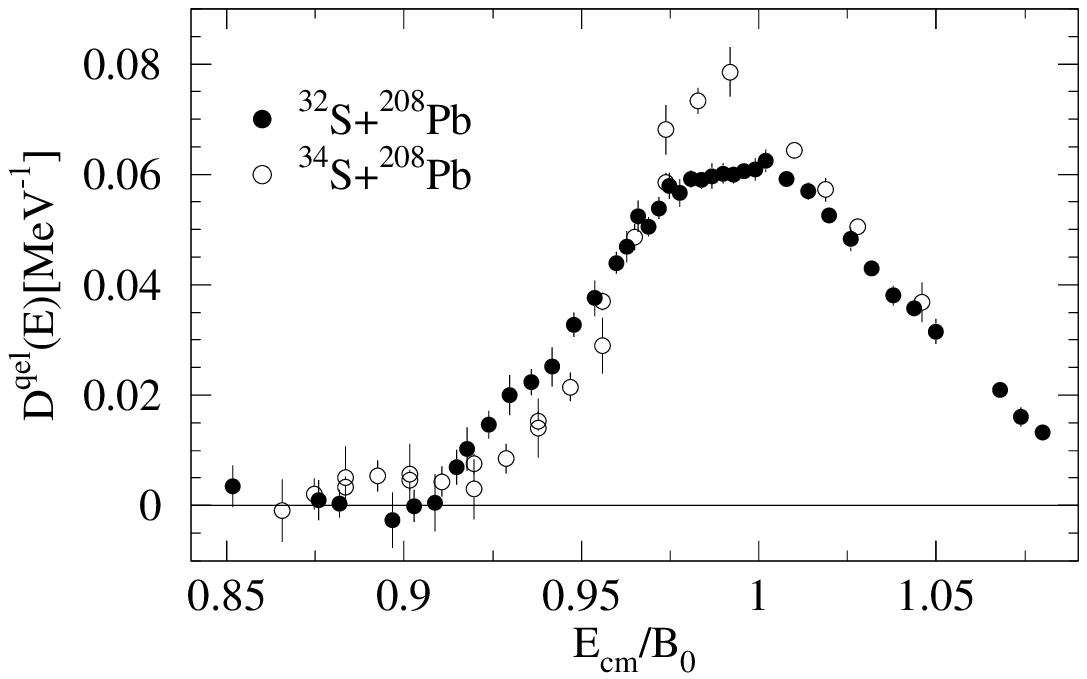}
\includegraphics[width=\linewidth]{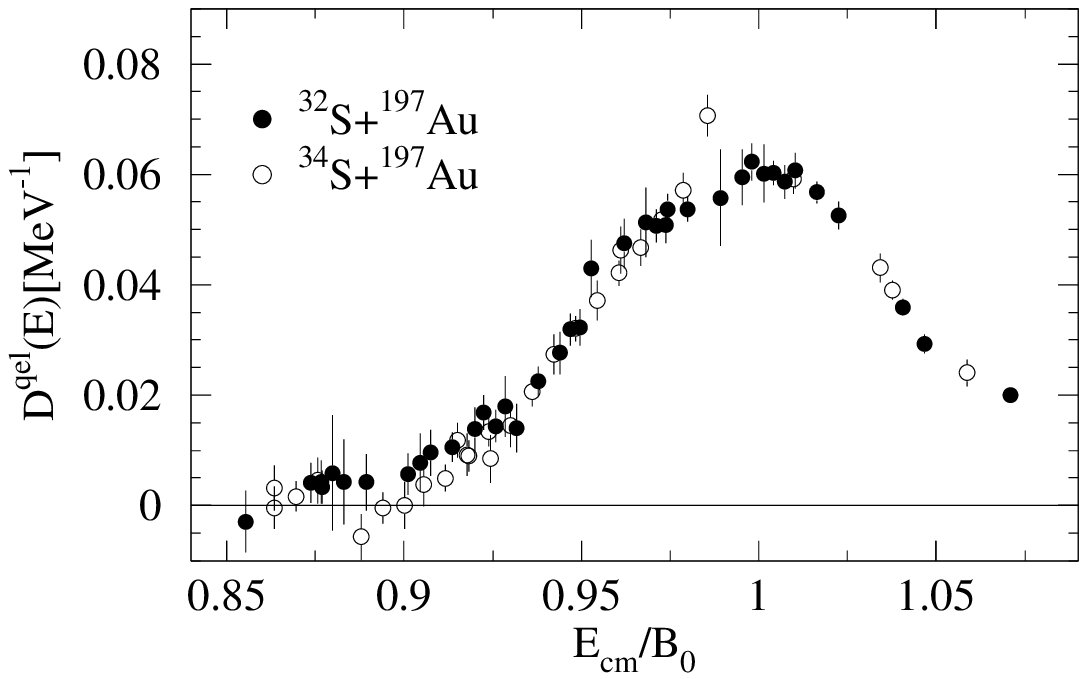}
\caption{Representations of the barrier distributions for \reaction{32,34}{S}{208}{Pb} (top) 
and  \reaction{32,34}{S}{197}{Au} (bottom). The energy scales have been normalised with the respective 
average barrier energy $B_0$.}
\label{allbd1}
\end{center}
\end{figure}

The barrier distribution representations $D^{qel}(E)$ have been extracted from the quasi-elastic scattering
 excitation functions $\frac{d\sigma^{qel}}{d\sigma^R}(E)$ by differentiation with respect to energy according to 
Equation~(\ref{dqel}). A
point-difference formula with discrete energy steps in the range $\Delta E_{lab} = $3--6\,MeV was used to evaluate 
the differential. The data sets obtained for 
the different energy steps are consistent, 
so that they have been combined. The resulting barrier 
distribution representations $D^{qel}(E)$ are shown for all four systems 
in Figure~\ref{allbd1}. In order to facilitate 
a direct comparison, the energy scales of the 
barrier distribution representations have been normalised by dividing by an average barrier $B_0$, which 
was chosen as the energy where  $\frac{d\sigma^{qel}}{d\sigma^{R}}(E) = 0.5$. 
Although this determination of the average barriers may seem somewhat arbitrary, the values obtained are
realistic. For example, 
the average barrier for the system \reaction{32}{S}{208}{Pb} has been 
determined as 144.4\,MeV by fitting the high energy part 
of the fusion excitation function using a single barrier penetration model~\cite{david}. 
This compares well with the value of $B_0$=144.2\,MeV used here.

As emphasized in the introduction, above the average barrier energy $B_0$ the representation $D^{qel}(E)$ is not sensitive 
to the fusion dynamics. Indeed, at these high energies $D^{qel}(E)$ is the same for 
all four systems. The low energy parts of the measured 
representations $D^{qel}(E)$ are discussed below.
 
For the system \reaction{34}{S}{208}{Pb} (open circles in Figure~\ref{allbd1} (top)) the slope of 
$D^{qel}(E)$ over the energy range $0.92~<~E/B_0~<~0.99$ is steeper than that 
for \reaction{32}{S}{208}{Pb} (filled circles in Figure~\ref{allbd1} (top)). Also, the maximum of $D^{qel}(E)$ 
for the reaction 
\reaction{34}{S}{208}{Pb} 
is 0.08 MeV$^{-1}$, whereas the maximum for the lighter system is only about 0.06 MeV$^{-1}$. 
Since the integral of $D(E)$ is unity, 
this implies
that the barrier distribution for \reaction{34}{S}{208}{Pb} 
is narrower than that for \reaction{32}{S}{208}{Pb}. This is consistent with significant coupling to 
positive Q-value neutron transfer channels 
in the \reaction{32}{S}{208}{Pb} fusion reaction. Indeed, both the one neutron ($Q = +1.3$~MeV) 
and two neutron transfer ($Q = +6.0$\,MeV)
Q-values for this system are positive. 

\begin{figure}
\begin{center}
\includegraphics[width=\linewidth]{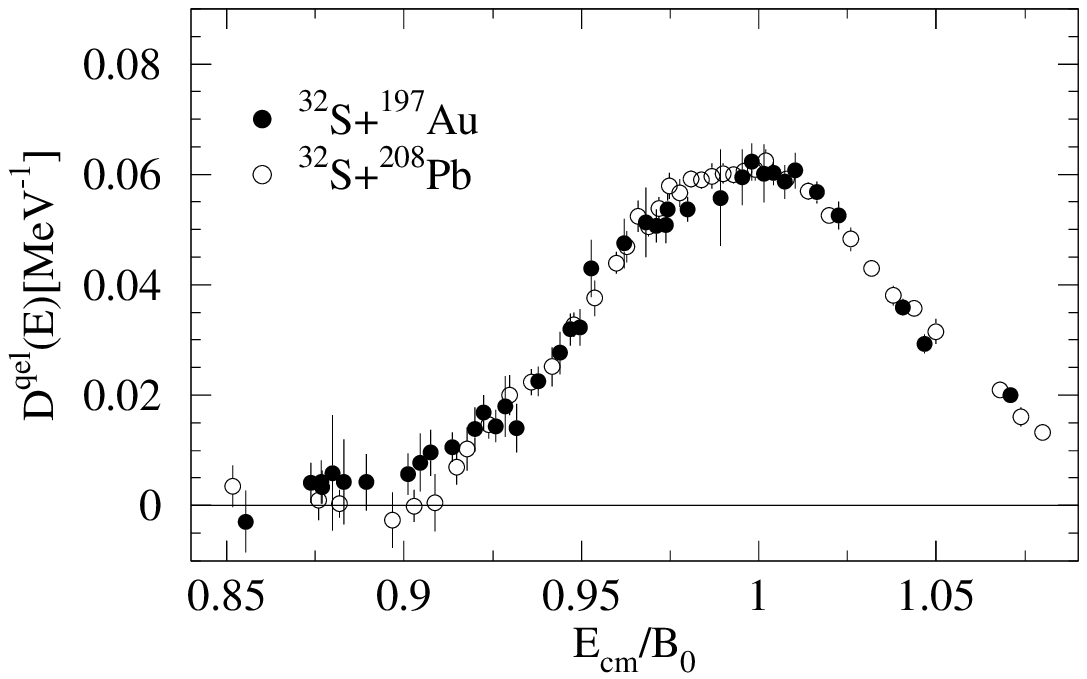}
\includegraphics[width=\linewidth]{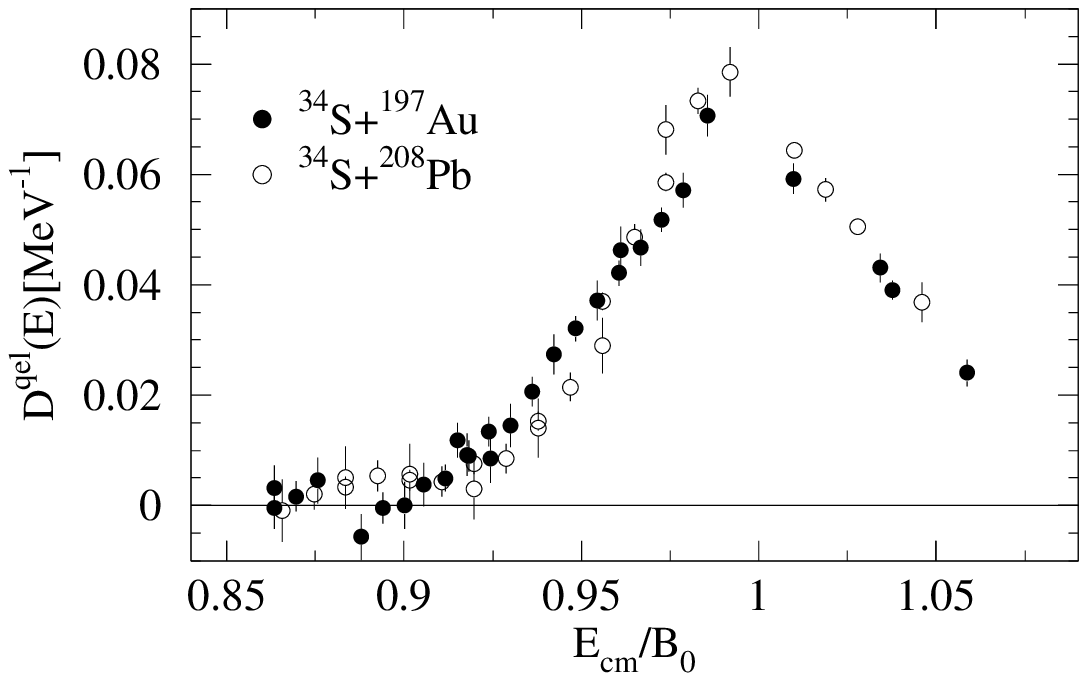}
\caption{Representations of the barrier distributions for \reaction{32}{S}{197}{Au},\iso{208}{Pb} (top) and 
\reaction{34}{S}{197}{Au},\iso{208}{Pb} (bottom).
 The energy scales have been normalised with the respective average barrier $B_0$.}
\label{allbd2}
\end{center}
\end{figure}

The equivalent data for the $^{197}$Au target (Figure~\ref{allbd1} (bottom)) do not show such a pronounced difference.
It is apparent from Figure~\ref{allbd2} (top) that the representations $D^{qel}(E)$ for  \reaction{32}{S}{197}{Au} and
\reaction{32}{S}{208}{Pb} agree, which would be consistent with these systems having similar barrier distribution and thus
equivalent coupling interactions. However, the comparison of the representations $D^{qel}(E)$ in Figure~\ref{allbd2} 
(bottom) for  \reaction{34}{S}{197}{Au} and \reaction{34}{S}{208}{Pb} demonstrates that for the gold systems 
$D^{qel}(E)$ is already broad for the heavier sulphur projectile, for which coupling to neutron transfer is less favoured. 
This suggests that coupling to states in the $^{197}$Au nucleus generates barrier strength at low energies, which is absent
for the reactions with $^{208}$Pb.

The experimental results for the two lead systems are consistent with 
those reported from fusion measurements 
for the two reactions \reaction{32,36}{S}{110}{Pd}~\cite{spd}, 
where additional barrier strength at low energies was
also found for the lighter projectile $^{32}$S. 
While the new data support an important role of positive Q-value neutron 
transfer channels in the fusion
of \reaction{32}{S}{208}{Pb}, such an interpretation is only unique, 
if the properties of the collective
states in $^{32}$S and $^{34}$S are identical, or at least can be assumed to be very similar. 
Recent results for the
fusion of the sulphur nuclei $^{32,34}$S with $^{89}$Y~\cite{Anjali} show 
that in that case the different collectivity of their quadrupole excitations (Table~2) 
results in a broader fusion barrier distribution for $^{32}$S than for $^{34}$S. 
Thus the differences observed in this work between the barrier distributions for $^{32}$S~+~$^{208}$Pb
and $^{34}$S~+~$^{208}$Pb may not be solely due to coupling to the positive Q-value
neutron pick-up channels. 
Measurements for the heavier system \reaction{36}{S}{208}{Pb}~\ may shed additional light on
the fusion mechanism. In this latter system the Q-values for 
both one neutron and two neutron transfer are negative (see Table \ref{qvalues}), so that
any effects due to neutron pick-up can be ruled out. 

\section{Conclusions}

The experiments reported here have demonstrated that precision measurements of 
quasi-elastic scattering at backward angles 
are able to probe the fusion barrier distribution of heavy systems below the average barrier.
Such measurements are thus in principle sensitive to the effects of positive Q-value transfer channels.
Indeed it was found that neutron transfer may affect the fusion of $^{32}$S with $^{208}$Pb. The results, however, are
also consistent with the observed additional barrier strengths at low energies being due to the increased collectivity of
$^{32}$S, when compared with $^{34}$S. For the fusion reactions of $^{32,34}$S with $^{197}$Au  
couplings related to the nuclear structure of $^{197}$Au appear to be dominant in shaping the low energy part
of the barrier distibution.

Since quasi-elastic scattering experiments are generally not as complex 
as fusion measurements, they are well suited to survey a number of reactions to determine
good candidates for detailed studies of the fusion dynamics. 
The extracted representations of the barrier distribution 
can be indicative of important coupling interactions, however, the conclusive identification of these couplings
may require the measurement and interpretation of the fusion excitation function.

\section*{Acknowledgements}
The authors are grateful to the late Prof. Trevor Ophel for his contributions 
to these experiments and would like to thank Dr David Hinde for indepth discussions of the results. The support 
of Dr Jack Leigh and
Dr Clyde Morton is also acknowledged.


\begin{thebibliography}{99}

\bibitem{reisdorf}
W.~Reisdorf, J. Phys. G 20 (1994) 1297, and references therein.

\bibitem{nandarev}
M.~Dasgupta, D.J.~Hinde, N. Rowley, and A.M. Stefanini, Ann. Rev. Nucl. Sci. 48 (1998) 401, and references therein.

\bibitem{dasso}
C.H.~Dasso, S.~Landowne, and A.~Winther, Nucl. Phys. A 405 (1983) 381; Nucl. Phys. A 407 (1983) 221.

\bibitem{rowley}
N.~Rowley, G.R.~Satchler, and P.H.~Stelson, Phys. Lett. B 254 (1991) 25.

\bibitem{qeldist}
H.~Timmers, J.R.~Leigh, M.~Dasgupta, D.J.~Hinde, R.C.~Lemmon, J.C.~Mein, C.R.~Morton, J.O.~Newton, and N.~Rowley, Nucl. Phys. A 584 (1995) 190.

\bibitem{capurro}
O.A.~Capurro, J.E~Testoni, D.~Abriola, D.E.~DiGregorio, G.V.~Mart\'\i, A.J.~Pacheco, and M.R.~Spinella, Phys. Rev. C 61 (2000) 037603;
O.A.~Capurro, J.E~Testoni, D.~Abriola, D.E.~DiGregorio, G.V.~Mart\'\i, A.J.~Pacheco, M.R.~Spinella, and E.~Achterberg, Phys. Rev. C 62 (2000) 014613.

\bibitem{santra}
S.~Santra, P.~Singh, S.~Kailas, Q.~Chatterjee, A.~Shrivasta, and K.~Mahata, Phys. Rev. C 64 (2001) 024602.

\bibitem{sinha}
S.~Sinha, M.R.~Pahlavani, R.~Varma, R.K.~Choudhury, B.K.~Nayak, and A.~Saxena, Phys. Rev. C 64 (2001) 024607.

\bibitem{cazr}
H.~Timmers, D.~Ackermann, S.~Beghini, L.~Corradi, J.H.~He, G.~Montagnoli, F.~Scarlassara, A.M.~Stefanini, and N.~Rowley, 
Nucl. Phys. A 633 (1998) 421.

\bibitem{hofmann}
S.~Hofmann, Rep. Prog. Phys., 61 (1998) 639.

\bibitem{heikophd}
H.~Timmers, Ph.D. thesis, Expressions of Inner Freedom, 
An Experimental Study of the Scattering and Fusion of 
Nuclei at Energies Spanning the Coulomb Barrier, 
http://thesis.anu.edu.au/public/adt-ANU20020328.152158, 
Australian National University, Canberra, Australia (1996).

\bibitem{erdadet}
H.~Timmers, T.R.~Ophel, and R.G.~Elliman, Nucl. Instr. Meth. Phys. Res. B 161-163 (2000) 19.

\bibitem{david}
D.J.~Hinde, A.C.~Berriman, R.D.~Butt, M.~Dasgupta, C.R.~Morton, J.O.~Newton, 
Dynamical Interplay of Fusion and Fission in
Low Energy Nucleus-Nucleus Collisions, Nucl. Phys. A 685 (2001) 72c-79c.

\bibitem{spd}
A.M.~Stefanini, D.~Ackermann, L.~Corradi, J.H.~He, G.~Montagnoli, S.~Beghini, F.~Scalarassa, and G.F.~Segato, Phys. Rev. C 52 (1995) 1727.

\bibitem{Anjali}
A.~Mukherjee, M.~Dasgupta, D.J.~Hinde, K.~Hagino, J.R.~Leigh, J.C.~Mein, C.R.~Morton, J.O.~Newton, and H.~Timmers, 
submitted to Phys. Rev. C.  
 
\end{thebibliography}
\end{document}